\documentclass[preprint]{elsarticle}

\usepackage{float}
\usepackage{fullpage}
\usepackage{array}
\usepackage{xcolor}
\usepackage{amsmath}
\usepackage{graphicx}
\graphicspath{ {./Figures/} }
\usepackage{caption}
\usepackage{subcaption}
\usepackage{bm}
\usepackage[utf8]{inputenc}
\usepackage[T1]{fontenc}

\newcommand{\myabstract}{
The driveability of a new heavy-truck driveline is traditionally assessed using physical prototypes. Enabling early evaluation of the driving experience in a human-in-the-loop driving simulator
using a virtual prototype has the potential to significantly improve development efficiency.

To enable driveability assessment using a moving-base simulator, participants must be able 
to perceive small differences in longitudinal acceleration.
The just-noticeable difference (JND) was therefore evaluated for two variants of the classical motion-cueing algorithm (MCA) tuned specifically for tip-in/launch tests and compared to a more general variant
in a driving simulator with a long linear track.
Psychometric functions were fitted to responses obtained using a weighted staircase procedure and analysed using a generalized linear model.

No significant differences in JND were found between the motion cueing variants.
The mean JND across all participants and MCA variants was 5.4\%.
The mean point of subjective equality in the JND experiment was -1.9\%, suggesting that participants perceived the acceleration as higher in the second stimulus of a pair.

In a subjective comparison, most participants preferred the motion cueing variants that were tuned for launch manoeuvres over
the general variant. 
}

\begin{document}

\begin{frontmatter}
\title{Effects of motion cueing on longitudinal acceleration perception in a driving simulator}
\date{\vspace{-5ex}}

\author[vti,isy]{Erik Gustaf Lilljebjörn\corref{cor1}}
\ead{erik.gustaf.lilljebjorn@vti.se}
\author[vti,isy]{Sogol Kharrazi}
\author[isy]{Jan Åslund}
\author[mai]{Martin Singull}

\address[vti]{Swedish National Road and Transport Research Institute (VTI)}
\address[isy]{Division of Vehicular Systems, Linköping University, Sweden}
\address[mai]{Division of Applied Mathematics, Linköping University, Sweden}

\cortext[cor1]{Corresponding author}

\begin{abstract}
\myabstract
\end{abstract}    

\end{frontmatter}

\section{Introduction}
The driveline-related driveability of a heavy truck is related to the subjective feeling of driving the truck.
It is desirable that the driving experience is comfortable and that the driveline feels responsive and strong.
Driveline-related driveability is commonly evaluated through a range of manoeuvres, with launch/tip-in from standstill being one of the main test cases \cite{Krenz1985,Kim2024,Dorsch2021}. 
A launch/tip-in is a highly informative test, as it excites the primary longitudinal drivetrain dynamics, including torque build-up, delay, jerk, and driveline oscillations.
Driveline-related driveability is typically evaluated in real vehicles by experienced test drivers using short tip-in/launch tests lasting up to five seconds.

Early in the development of a heavy-truck driveline, 
physical prototypes are typically not available for testing. At the same time, it is costly to correct poor design choices late in the project.
Consequently, early evaluation of driveability using a virtual prototype in a moving-base driving simulator can significantly benefit truck manufacturers.

In a moving-base driving simulator, in addition to visual and auditory cues, motion feedback is used to give the driver a realistic driving experience. 
Because the workspace of a moving-base driving simulator is limited, a motion-cueing algorithm (MCA) maps the motions of the vehicle model to feasible platform motions.
The classical MCA splits the vehicle acceleration into a low and a high frequency part using low-pass and high-pass filters \cite{Reid1985}. The low frequency part is rendered by tilting the cabin,
whereas the high frequency part is recreated using the translational acceleration of the simulator platform.  

There are few studies that assess longitudinal driveability in a moving-base driving simulator, and the majority generally does not investigate how motion cueing affects the driveability assessment.
For example, Erler et al. \cite{ErlerShuffle} investigated driveline shuffle following a tip-in/back-out manoeuvre. A driving simulator with a 10-meter longitudinal track was used, 
and the tip-in phase lasted less than 2 seconds. As a result, vehicle motions could be rendered
using only translational acceleration of the cabin, without the need for acceleration scaling or cabin tilt.
In another study, Kraft et al. \cite{KraftGearShifts} showed that the subjective perception of gear shifts can be evaluated in a moving-base driving simulator. Because participants were not driving
actively, the motion cues could be optimised offline in advance, and the study therefore provides no guidance on how to tune the MCA.

Baumgartner et al. \cite{Baumgartner2} investigated how the scale factor of the MCA affects the just-noticeable difference (JND) in acceleration. The JND in acceleration is relevant for driveability assessment
because changes in acceleration that can be perceived in the real vehicle should also be detectable in a driving simulator.

To the authors' knowledge, it has not been studied previously how the MCA tuning affects subjective driveability assessments in a driving simulator. The aim of the present study is therefore to investigate how different
variants of the classical MCA are perceived by professional test drivers in a short tip-in manoeuvre. The authors are not aware of any previous studies examining the effect of MCA cut-off frequencies on acceleration perception. 
The present study therefore aims to investigate their influence on the JND in acceleration.  

Three variants of the classical MCA will be compared. 
Two of the variants are tuned specifically for short tip-in tests and one is a more general variant that can be used in a wide range of scenarios.

\subsection{JND estimation}
The just-noticeable difference (JND) is the smallest difference in the intensity of a stimulus that a person can perceive. JND in acceleration is usually estimated using a two-alternative forced-choice (2AFC) method
in which a standard and a comparison stimulus are presented successively to the test subject.
After each such trial, the subject is asked if the comparison was lower or higher than the standard stimulus.

\begin{figure}[H]
    \centering
    \includegraphics[width=0.5\textwidth]{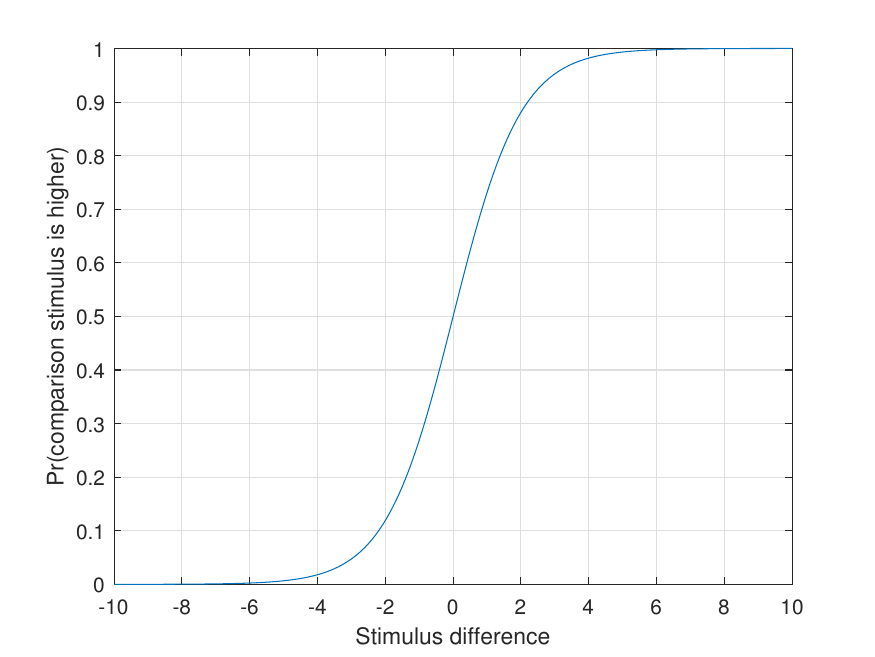}
    \caption{Psychometric function relating the probability of a binary outcome to stimulus difference (arbitrary units).}
    \label{fig:GenericPsychometric} 
\end{figure}

In psychophysical experiments, a psychometric function, such as the one shown in Figure \ref{fig:GenericPsychometric}, relates the probability of a binary outcome 
(e.g., a correct response or the comparison stimulus exceeding the standard) to a stimulus variable, either the stimulus intensity in a detection task or the difference in stimulus intensity in a discrimination task.
In the remainder of this section, the stimulus variable is expressed as a stimulus difference, but the methods also apply directly to stimulus intensity in detection tasks.
A psychometric function can be estimated using the method of constant stimuli, in which a large number of trials are performed at a number of fixed stimulus differences \cite{Gescheider2013}.
If it is sufficient to estimate a single threshold on the psychometric function, which is often the case, an adaptive test procedure can be used to estimate the threshold 
efficiently with fewer trials than with a fixed-level method \cite{Leek2001}. A common adaptive procedure is the up/down staircase in which the stimulus difference is increased following an incorrect answer,
or decreased following one or more correct answers. The simple up/down staircase uses equal step sizes in both directions, and a single correct answer is required to take a step down. It is used in yes/no detection tasks
to estimate the stimulus intensity at which the probability that the stimulus is detected is 50\%.

To estimate a threshold corresponding to any other probability, e.g., the 75\% point, in a 2AFC task, a weighted staircase, in which the step size up and the step size down are different, can be used \cite{Kaernbach1991}.
The step sizes are chosen such that the expected change in the stimulus difference, \(E(\Delta)\), is zero at the target level.
The magnitude of the change in stimulus difference, \(\Delta\), is either equal to \(\Delta^-\), following a correct response, or equal to \(\Delta^+\), following an incorrect response.
With a probability \(p^*\) of a correct response at the unknown target threshold, we have
\[
E(\Delta) = \Delta^+ (1-p^*) -\Delta^- p^* = 0
\]
\begin{equation} \label{eq:Stepsize}
    \Leftrightarrow \frac{\Delta^+}{\Delta^-} = \frac{p^*}{1-p^*}.
\end{equation}
Because \(E(\Delta)\) is equal to zero at the target threshold, \(E(\Delta) > 0\) at stimulus differences where \(p < p^*\), and \(E(\Delta) < 0\) at stimulus differences where \(p > p^*\),
the weighted up/down staircase will reach the target level and then oscillate around that level.

The JND in acceleration has been investigated on the road in a real car and in moving-base simulators in several previous studies. 
Müller et al. \cite{Muller2013} investigated the JND in acceleration during a full-load tip-in manoeuvre in a real car. 
An important contribution of their work is the recognition that perceptual thresholds in acceleration and jerk are highly relevant for driveline development, and that such thresholds can be studied in a realistic driving
scenario. However, a simple up/down staircase procedure was used that yields stimulus differences around the level corresponding to a 50\% probability of a correct response, i.e., chance performance in a 2AFC task.
Consequently, the JND estimates obtained in \cite{Muller2013} may not represent the level at which participants can reliably detect differences between stimuli.

Several other studies investigated acceleration JNDs in a controlled environment using moving-base simulators.
Naseri et al. \cite{Naseri2012} measured acceleration JNDs in a flight simulator using sinusoidal acceleration profiles, reporting values ranging from 0.05 m/s\textsuperscript{2} 
at a reference peak acceleration of 0.5 m/s\textsuperscript{2} to 0.13 m/s\textsuperscript{2} at 2.0 m/s\textsuperscript{2}. They also found JND estimates to increase with increasing frequency.

\begin{figure}[H]
    \centering
    \begin{subfigure}[b]{0.4\textwidth}
        \centering
        \includegraphics[width=\textwidth]{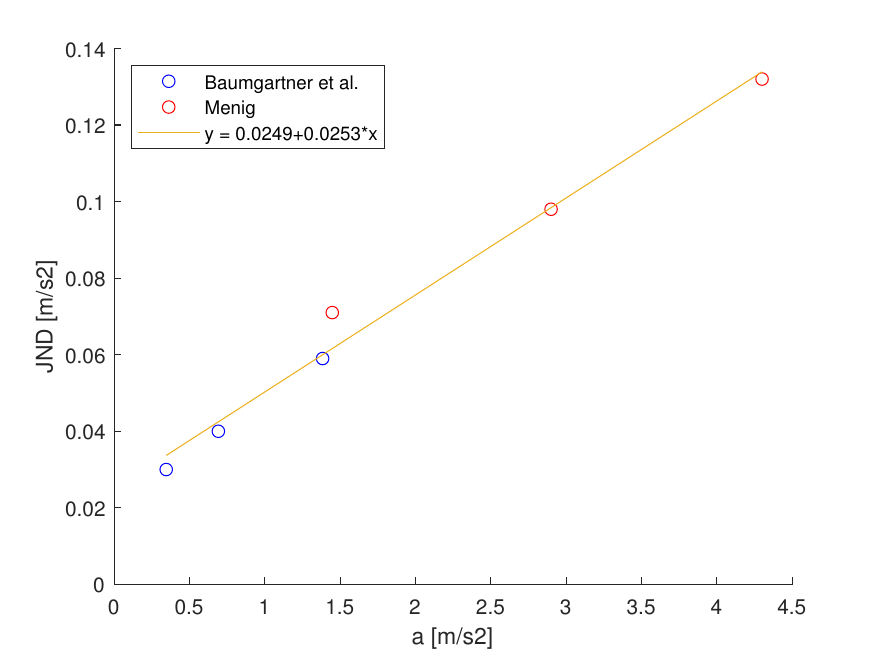}
        \caption{Acceleration JND vs reference acceleration.}
        \label{fig:JndVsAccBaumMenig}
    \end{subfigure}
    \begin{subfigure}[b]{0.4\textwidth}
        \centering
        \includegraphics[width=\textwidth]{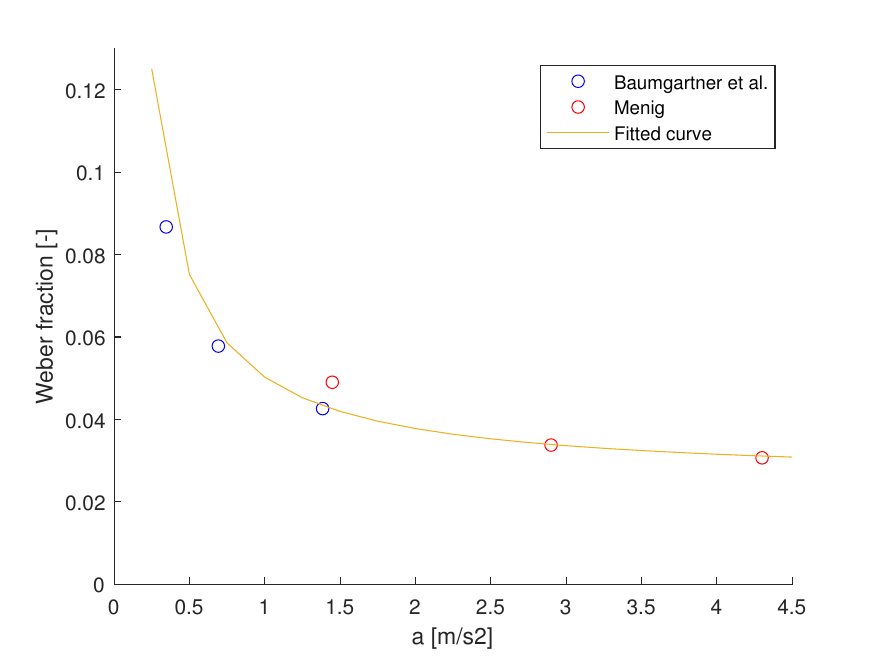}
        \caption{Weber fraction vs reference acceleration.}
        \label{fig:WeberFracBaumMenig}
    \end{subfigure}
    \caption{Results recreated from Baumgartner et al. \cite{Baumgartner2} and Menig \cite{Menig2025}.}
    \label{fig:JndWeberBaumMenig} 
\end{figure}

Using a hexapod-based driving simulator, Baumgartner et al. \cite{Baumgartner2} reported mean JNDs in acceleration ranging from 0.03 m/s\textsuperscript{2} at a reference acceleration of 0.346 m/s\textsuperscript{2} to 0.059 m/s\textsuperscript{2} 
at 1.384 m/s\textsuperscript{2}, in a full-load tip-in manoeuvre. Further, they showed that the JND increases approximately linearly with the reference acceleration, as seen in Figure \ref{fig:JndVsAccBaumMenig}. 
Consistent with classical psychophysical theory \cite{Gescheider2013}, the ratio between the JND and the reference stimulus intensity, known as the Weber fraction, 
was shown to be approximately constant for reference accelerations of 1.5 m/s\textsuperscript{2} and above, but increases rapidly near the absolute detection threshold, as seen in Figure~\ref{fig:WeberFracBaumMenig}.
Because the Weber fraction is approximately constant in this range, a driving simulator can be used to investigate whether driveline changes resulting in an increase or decrease in acceleration are noticeable to drivers, provided that 
the scaled-down accelerations in the simulator are within this range.

In a driving simulator equipped with a 10-meter-long track, and using a similar tip-in manoeuvre, Menig \cite{Menig2025} observed acceleration JNDs ranging from 0.071 m/s\textsuperscript{2} at 
a 1.448 m/s\textsuperscript{2} reference acceleration to 0.132 m/s\textsuperscript{2} at 4.3 m/s\textsuperscript{2}. In Figure \ref{fig:JndWeberBaumMenig}, the results reported by Menig \cite{Menig2025} agree well with those observed by 
Baumgartner et al. \cite{Baumgartner2}.

\section{Driving simulator}
\begin{figure}[H]
    \centering
    \includegraphics[width=0.5\textwidth]{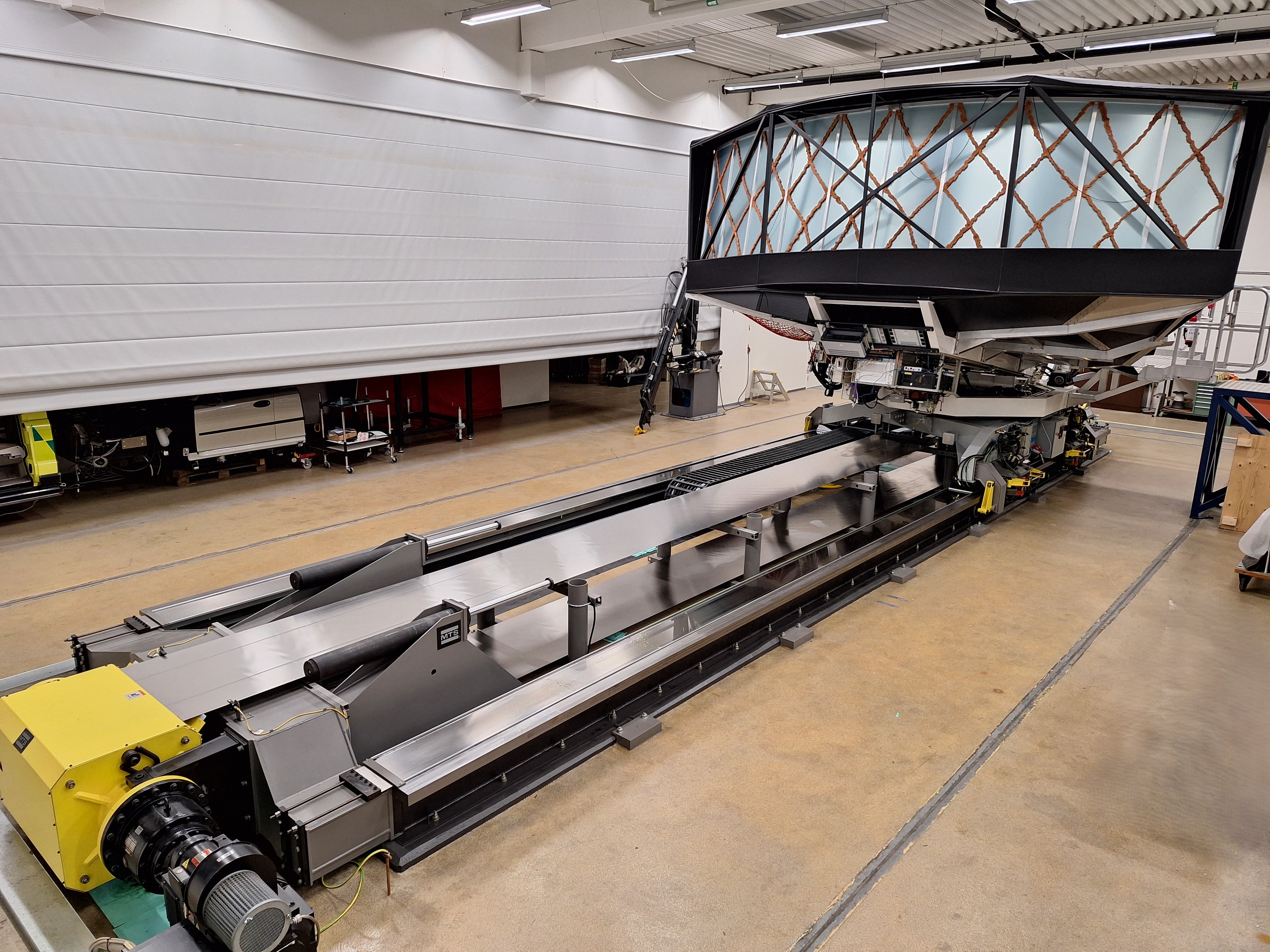}
    \caption{Driving simulator used in the study.}
    \label{fig:SimIII}
\end{figure}

The driving simulator used in the study, shown in Figure \ref{fig:SimIII}, is equipped with a separate system for rendering translational accelerations using a long track, unlike hexapod-based simulators that use
the same actuators for all motions (e.g., surge and pitch). The track can be used to render surge or,
if the platform is rotated 90 degrees, sway motion.
The maximum travel, speed and acceleration of the translational motion are 7.5 m, 3.75 m/s, and 8.0 m/s\textsuperscript{2}, respectively.
In addition, the simulator cabin can be tilted in pitch and roll to simulate sustained longitudinal and lateral accelerations of up to 1.5 m/s\textsuperscript{2}. The simulator is equipped with
a vibration table that is used to generate vibrational cues and to reproduce vehicle roll and pitch dynamics.

Normally, the simulator starts in the center of the longitudinal track and returns to this position when the vehicle is not accelerating.
Since the tip-in manoeuvre does not involve hard braking, the simulator can be pre-positioned near the start of the track to maximize the available translational workspace.

\subsection{Vehicle model}
The vehicle model is a 6-DOF multi-body dynamics model of a 6x4 tractor and semi-trailer. A simple model of an electric driveline that can capture
the main behaviour of the real truck driveline in a full-throttle tip-in test was developed for the study.
The simple model consists of a max torque-motor speed curve and a single total gear ratio from the motor shaft to the wheels. The dynamics of the driveline is modelled
as a 2nd order low-pass filter that was tuned to get similar maximum jerk in the simulator as in the real truck. 
The maximum acceleration can be controlled by changing the maximal torque.

\section{Motion cueing}
In a moving-base driving simulator, the goal of motion cueing is to give the driver a convincing experience of being in a moving vehicle, even though the  
motions of the simulator itself are very restricted.
To achieve this, the simulator uses information from the vehicle model, such as translational accelerations and angular velocities, to create motion cues that are presented to the driver using
the moving base.
There are essentially two ways that the motion cueing can render translational acceleration; using actual translational acceleration, or by tilting the cabin.
Because the workspace of the simulator is limited, only quick changes in acceleration, such as sudden starts and stops, can be recreated accurately using
actual translational acceleration. Sustained accelerations, on the other hand, can be rendered by slowly tilting the cabin. Provided that the cabin is tilted slowly and without the help of visual cues, 
the human brain is unable to distinguish cabin tilt from translational acceleration. The drawback of using cabin tilt is that, unless the angular velocity and acceleration
is kept low, participants may notice that they are being rotated and not accelerated, which breaks the illusion.

Tuning the motion cueing involves some interesting and difficult trade-offs. For instance, one has to decide what is more
important, that the amplitude of the perceived acceleration is close to that of 
the actual vehicle acceleration or that the perceived acceleration has the correct shape and that false motion cues are avoided. 
Also, one has to decide if it is important to keep the cabin angular speed below the perception threshold.
If speeds above the thresholds can be accepted then higher scale factors can be used without distorting the shape of the acceleration.

How one decides to tune the motion cueing of course depend on the application.
It is reasonable to think that for driveability studies, the motion cueing should be tuned in such a way that false cues are avoided.
It is possible that missing or false motion cues can be perceived as deficient driveability which makes driveability assessment impossible. 
Therefore, accelerations should be scaled down as much as needed to avoid disturbing false cues. 
But on the other hand, if the acceleration is scaled down so that the rendered acceleration is very low, the driver might not be able to perceive real
driveability issues.

\begin{figure}[H]
    \centering
    \includegraphics[width=0.5\linewidth]{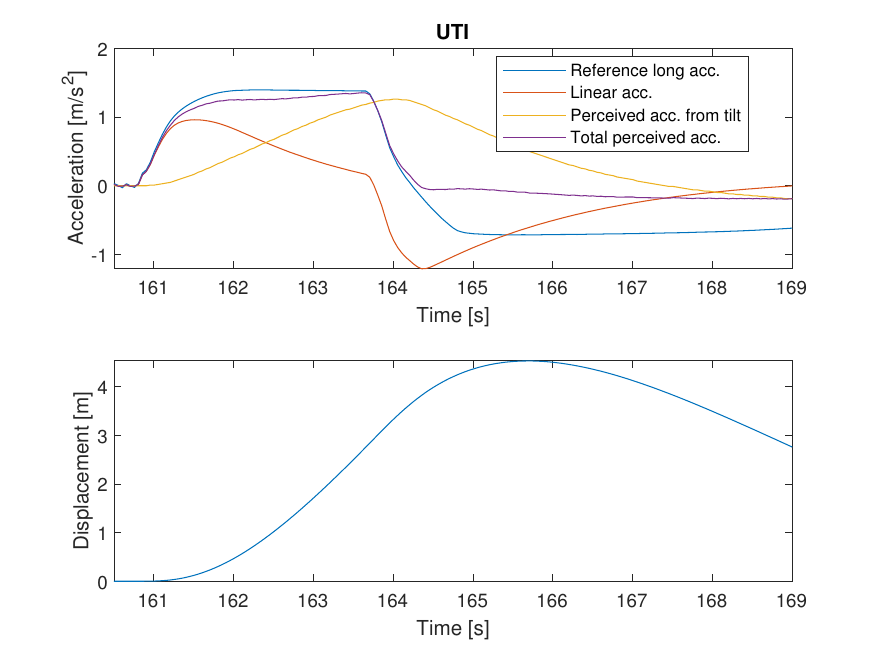}
    \caption{Platform acceleration and displacement and cabin tilt of motion cueing UTI.}
    \label{fig:UTI outputs}
\end{figure}

The classical MCA was used in the study because it is widely used and is easy to tune. Also, it was already implemented in the simulator before the study.
Because the translational channel uses a 2nd order high-pass filter, the simulator is not returned to the starting position as long as there is a non-zero acceleration, as shown in
the bottom subplot of Figure \ref{fig:UTI outputs}.
The low-pass and high-pass filters are complementary, i.e., the sum of the gains of the filters is equal to one for any given frequency.
The cabin tilt speed is limited to 3 deg/s in all variants, which is considered to be below the perception threshold of most people \cite{Groen2004}.
The rotational acceleration is limited to 8 deg/s\textsuperscript{2}.

\begin{figure}[H]
    \centering
    \includegraphics[width=0.5\linewidth]{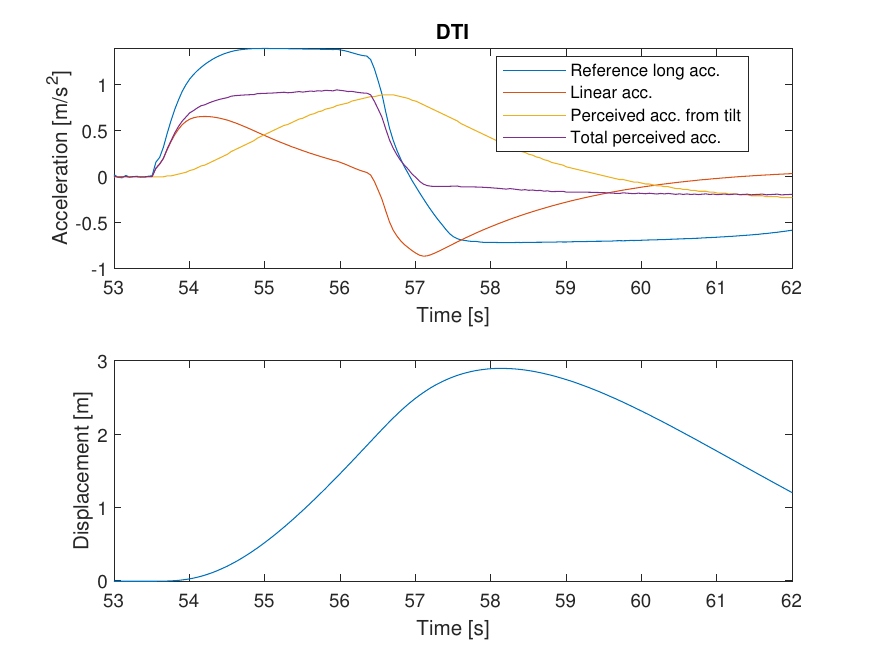}
    \caption{Platform acceleration and displacement and cabin tilt of motion cueing DTI.}
    \label{fig:DTI outputs}
\end{figure}

\begin{figure}[H]
    \centering
    \includegraphics[width=0.5\linewidth]{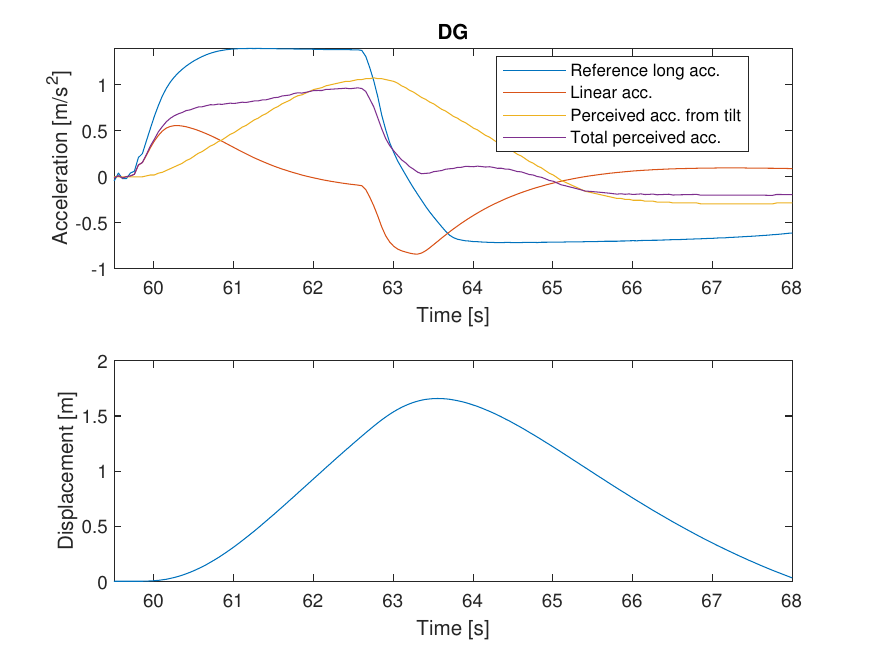}
    \caption{Platform acceleration and displacement and cabin tilt of motion cueing DG.}
    \label{fig:DG outputs}
\end{figure}

In preliminary tests, it was found that participants experience much less simulator sickness if the simulator is returned to the starting position with low deceleration.
The deceleration was therefore limited to 0.2 m/s\textsuperscript{2} in the study and the magnitude of the perceived specific force is consequently much lower than the reference acceleration 
when the vehicle is braking in Figures \ref{fig:UTI outputs}-\ref{fig:DG outputs}.

The MCA variants considered in the study are the Unscaled Tip-In (UTI), Downscaled Tip-In (DTI), and Downscaled Generic (DG) variants.
The UTI variant was tuned specifically for short tip-in manoeuvres, the acceleration is not scaled down, and it uses the translational motion extensively. 
The cut-off frequency is tuned so that the simulator uses the full length of the linear rail in a three seconds long full-throttle 
tip-in manoeuvre with 1.4 m/s\textsuperscript{2} acceleration when the simulator is pre-positioned near the start of the rail before the manoeuvre.
The DTI variant was also, like UTI, tuned specifically for short tip-in tests but the acceleration is scaled down. 
Tip-in manoeuvres with a peak acceleration of 1.4 m/s\textsuperscript{2} can be recreated for up to five seconds using DTI before the simulator reaches the end of the rail.
The DG variant uses cabin tilt extensively and can recreate a constant 1.4 m/s\textsuperscript{2} acceleration indefinitely without exceeding the workspace limits, making it the most general of the considered variants. 
The acceleration is scaled down and the cut-off frequency is tuned based on the steady-state response of the simulator to a
1.4 m/s\textsuperscript{2} step input.
The properties of motion cueing UTI, DTI and DG are summarized in Table \ref{table:MCAs}.

\begin{table}[H]
\begin{center}
    \begin{tabular}{>{\raggedright\arraybackslash}p{0.1\textwidth} >{\raggedright\arraybackslash}p{0.1\textwidth}
        >{\raggedright\arraybackslash}p{0.1\textwidth} >{\raggedright\arraybackslash}p{0.1\textwidth} >{\raggedright\arraybackslash}p{0.3\textwidth}}
    \hline
    Motion cueing & Scale factor & Cut-off frequency [rad/s] & Max tilt speed [deg/s] & Comment \\ 
    \hline
    1: UTI & 1 & 0.3 & 3 & Tuned specifically for short tip-in tests. \\
    \hline
    2: DTI & 0.7 & 0.32 & 3 & Tuned specifically for tip-in tests but is more general than UTI.  \\
    \hline
    3: DG & 0.7 & 0.55 & 3 & Uses cabin tilt extensively and is the most general motion cueing.\\ 
    \hline
\end{tabular}
\caption{Summary of the motion cueing variants}
\label{table:MCAs}
\end{center}
\end{table}

\section{Study design}
The study consists of two experiments. The first experiment aims to determine how the motion cueing variants affect JNDs in acceleration.
The second experiment is a pairwise comparison of the motion cueing variants where the participants were asked to rate the variants subjectively. The test procedures of the two experiments
are described in this section.

\subsection{Driving manoeuvre}
The same driving manoeuvre was used in both experiments. A pre-defined manoeuvre is used in the study so that the motion cueing can use the full length of the linear rail. 
The manoeuvre is a full-throttle tip-in from standstill
with a maximum acceleration of 1.4 m/s\textsuperscript{2}. 
A reference acceleration below the 1.5 m/s\textsuperscript{2} maximum sustained acceleration was chosen so that
the acceleration can be increased in the JND experiment.

The pre-defined tip-in starts when the driver presses the accelerator pedal more than 50\%. The participants were instructed to keep the pedal pressed down during the manoeuvre 
as they normally would during a tip-in test. After around three seconds, the tip-in ends automatically and the simulator gently returns to the
starting position. The reference acceleration in the tip-in manoeuvre can be seen in Figures \ref{fig:UTI outputs}-\ref{fig:DG outputs}.

The vehicle is driven on a straight highway with no other vehicles on the road, so the test persons can concentrate on the feeling of the longitudinal acceleration.

\subsection{Just-noticeable difference experiment}
In this experiment, the stimulus is the longitudinal acceleration during a full-throttle tip-in.
Each participant completed 20 trials per motion cueing variant. A trial consists of a pair of tip-in/drive-off manoeuvres; first the reference acceleration profile (the standard stimulus) 
and then a second tip-in where the acceleration had been slightly lowered or increased (the comparison stimulus). The direction of the change was quasi-random.
For practical reasons, only five absolute difference levels were used (4\%, 6\%, 8\%, 10\% and 12\%).   

To place trials appropriately on the stimulus difference axis, a weighted up-down test procedure with a step size up twice as large as the step size down was used in the present study.
It follows from Equation \eqref{eq:Stepsize} that if \(\Delta^+/\Delta^- = 2\) then \(p = 2/3\).

After 10 trails, the test persons rated their motion sickness using the Fast Motion Sickness (FMS) scale, a 20-point scale ranging from 0 (no sickness at all) to 20 (frank sickness) \cite{Keshavarz2011}. 

\subsection{Subjective comparison}
This experiment is a pairwise comparison of the motion cueing variants. Each variant is compared to every other variant (e.g., UTI-DTI, UTI-DG, DTI-DG),
resulting in a total of three comparisons per participant.
The drivers first tested the first motion cueing variant in a pair by performing a series of full-throttle tip-in manoeuvres. The simulator was then parked so that the motion cueing could be changed to the 
second variant and then the test person performed the same manoeuvre with the second variant. The participants were allowed to do as many full-throttle tip-in manoeuvres 
as they needed with a motion cueing variant to get a lasting impression of each variant. They usually tested a variant around 5-10 times. Finally, the test persons were asked to rate if the second 
variant in the pair was much worse, worse, slightly worse, the same, slightly better, better or much better than the first variant. The experiment was balanced so that all variants were presented first and last
in a comparison an equal number of times.

\subsection{Participants}
The study included 10 participants. All participants were driveability experts at a heavy truck manufacturer with prior experience of battery-electric heavy trucks similar to the one modelled in the study.

\section{Results and discussion}
The results of the JND experiment and the subjective comparison are presented and discussed in this section.
\subsection{Just noticeable difference}
In this section, the effects of the motion cueing variants on the JND are analysed. One participant was unable to complete the study due to simulator sickness.
None of the nine participants who completed the study reported motion sickness symptoms greater than 7 on the FMS scale at any time during the JND experiment.

\begin{figure}[H]
    \centering
    \begin{subfigure}[b]{0.48\textwidth}
        \centering
        \includegraphics[width=\textwidth]{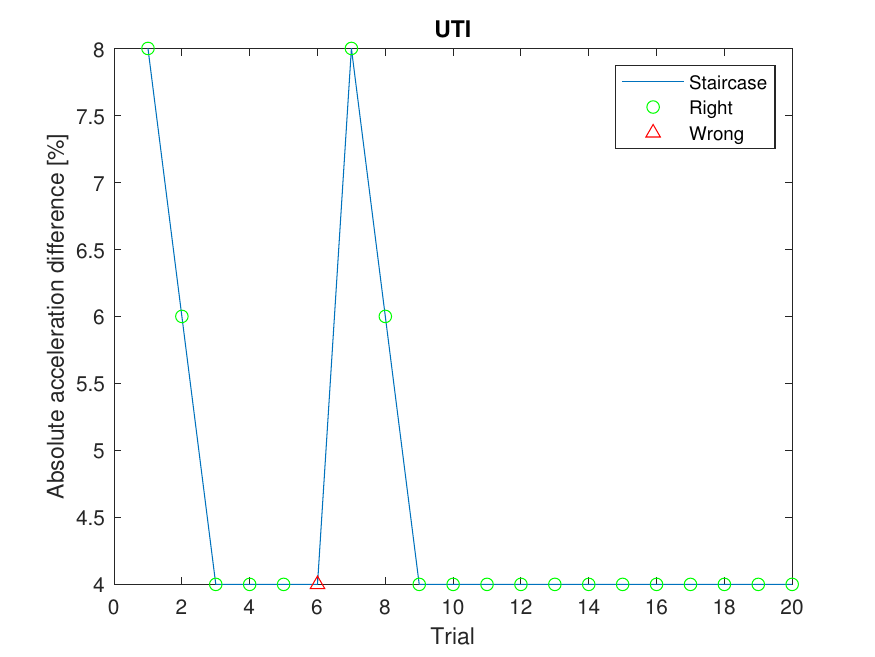}
        \caption{Example 1}
        \label{fig:Low_JND_staircase}
    \end{subfigure}
    \begin{subfigure}[b]{0.48\textwidth}
        \centering
        \includegraphics[width=\textwidth]{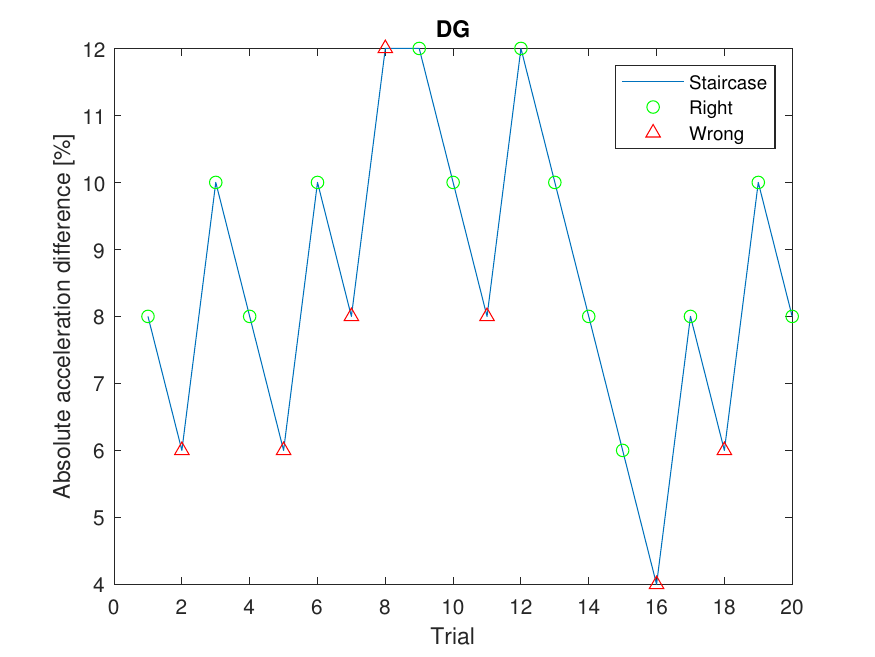}
        \caption{Example 2}
        \label{fig:High_JND_staircase}
    \end{subfigure}
    \caption{Example staircases of trials.}
    \label{fig:Staircases}
\end{figure}

Two adaptive staircases from the JND experiment are shown in Figure \ref{fig:Staircases}. In Figure \ref{fig:High_JND_staircase}, the test person
is unable to answer correctly most of the time when the difference in acceleration is small (6\% or less) and therefore the
acceleration difference is around 8-10\% in most trials. In Figure \ref{fig:Low_JND_staircase},
the test person is able to answer correctly even at the 4\% level most of the time.

With an adaptive staircase method, a threshold on the psychometric curve is usually estimated by calculating the mean of the 
reversals (peaks and valleys) \cite{Wetherill1966}. In the present study, for a number of reasons (outlined below), the JND was instead estimated using a 
hybrid method \cite{hall1981hybrid}, in which stimulus intensities were chosen using a staircase procedure, and a psychometric function was fitted to the responses
in the 2AFC task post hoc using a binary generalized linear model, that will be described in the following section, to estimate the JND.

The chosen up/down rule in the experiment targets \(X_{0.67}\) but with the hybrid method, another threshold, for instance \(X_{0.75}\), can be estimated. 
Also, with an up/down procedure, stimulus intensities are supposed to oscillate up and down around the target threshold. 
The difference in acceleration was limited to be at least 4\% for practical reasons. In Figure \ref{fig:Low_JND_staircase}, 
it is clear that for this participant, the target threshold is below 4\% and the procedure will not oscillate around the target threshold.
Finally, García-Pérez \cite{Garcia} showed that in small-sample experiments (less than 20 reversals), 
the staircase estimates will be strongly biased by the initial stimulus intensity in the experiment. 

\subsubsection{Effects of the motion cueing variants on the JND}
\begin{figure}[H]
    \centering
    \includegraphics[width=0.5\textwidth]{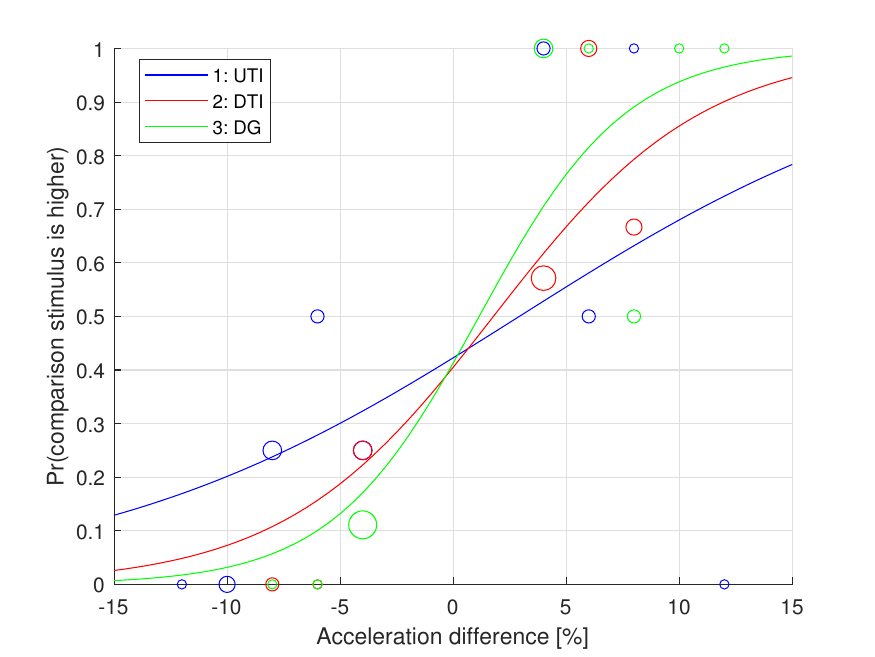}
    \caption{Fitted psychometric functions for the three MCA variants for one of the drivers.}
    \label{fig:Fitted PFs}
\end{figure}

The binary responses, \(y_i\), in the 2AFC task ("the comparison stimulus was lower/higher than the standard stimulus") are Bernoulli random variables,
with probabilities \(p_i\) for \(y_i = 1\) (comparison stimulus was higher). 
The expected value of the response in the i'th trial is \(\mu_i = E(y_i) = p_i\).
The probability that a subject answers that the comparison stimulus was higher than the standard stimulus can be modelled as a generalized linear model
that relates the expected value of the response variable to a linear function of the explanatory variables via a non-linear link function, \(g(\mu)\). In our analysis,
the explanatory variable is the difference in acceleration between the comparison and standard stimuli, and the link function is the 
logit function, \(g(\mu) = log(p/(1-p))\). With this link function, the probability is bounded to be between 0 and 1, and we have
\begin{equation} \label{eq:GLM}
    g(\mu_i) = log \left( \frac{p_i}{1-p_i} \right) = \beta_0 + \beta_1 x_i,
\end{equation}
where \(x_i\) is the acceleration difference and \(\beta_0\) and \(\beta_1\) are unknown parameters.
The probability that a subject answers that the comparison acceleration is higher is then the inverse of the link function, the logistic function 
\begin{equation} \label{eq:PF}
    p(x_i) = \frac{1}{1+e^{-(\beta_0+\beta_1 x_i)}}.
\end{equation}
The psychometric functions were fitted using maximum-likelihood estimation with the \textit{fitglm} function in MATLAB's Statistics and Machine Learning toolbox.
The point on the psychometric function where \(\text{Pr(\textit{comparison stimulus is higher}}) = 0.5\) is known as the point of subjective equality (PSE). 
At this stimulus difference, a test person perceives the standard and comparison stimuli as being equally strong.
An upper JND for positive differences can be calculated as the difference between the 0.75 point and the PSE and a lower JND for negative differences is calculated as the difference between the PSE and the 0.25 point \cite{Gescheider2013}. 
The upper and lower JNDs are averaged to obtain a single JND value for each participant and MCA variant.
An example of fitted psychometric functions for one of the drivers is shown in Figure \ref{fig:Fitted PFs}.
The JND estimates for each participant and MCA variant are shown in Table \ref{tbl:JND estimates}.

\begin{table}[h!]
\centering
\begin{tabular}{llllllllllll}
    \hline
    & 1 & 2 & 3 & 4 & 5 & 6 & 7 & 8 & 9 & Mean & Std \\
    UTI & 4.1 & 2.5 & 2.3 & 2.0 & 14 & 10 & 11 & 3.7 & 3.9 & 5.9 & 4.4 \\ 
    DTI & 6.9 & 6.4 & 1.5 & 3.5 & 9.4 & 5.1 & 6.3 & 5.5 & 7.4 & 5.8 & 2.3 \\ 
    DG & 12 & 0.92 & 2.6 & 3.8 & 1.9 & 3.6 & 5.1 & 3.8 & 24 & 6.4 & 7.2 \\ 
    \hline 
\end{tabular}
\caption{JND estimates of test subjects 1-9 for each MCA variant, rounded to two significant digits.}
\label{tbl:JND estimates}
\end{table}
Since the experiment employed a within-subjects design, paired t-tests were used to test if the mean JNDs differ between any of the MCA
variants that scale down the acceleration (DTI and DG) and the UTI variant, which does not. For both comparisons, the null hypothesis was that the mean JNDs of the two variants are equal.
The paired t-test assumes normally distributed data. Although the distribution of the individual JND estimates were right-skewed, the within-participant differences were approximately normally distributed.
Kolmogorov-Smirnov tests confirmed that the differences in JND between DTI and UTI, as well as between DG and UTI, did not significantly deviate from normality (p = 0.63 and p = 0.72, respectively). 

According to classical psychophysical theory, the ratio of the JND to the reference stimulus, known as the Weber fraction, is approximately constant for a wide range of stimulus intensities, but increases at small 
intensities close to the absolute detection threshold \cite{Gescheider2013}. Consequently, we expected the mean JNDs with the DTI and DG variants that scale down the acceleration to be equal to or greater than the mean JND
with the UTI variant. Because the direction of the effect of the scaling is known, one-tailed t-tests were used. 

The mean difference in JND between the DTI and UTI variants is \(\overline{\Delta JND} = -0.165\) and the sample variance is \(s^2 = 13.6\) so the t-statistic is
\[
t = \frac{\overline{\Delta JND}}{\sqrt{s^2/N}} = -0.13
\]
with a p-value of 0.55. Since \(p = 0.55 > \alpha = 0.05\), the null hypothesis cannot be rejected. The mean JNDs of the UTI and DTI variants are similar (see Table \ref{tbl:JND estimates}), so this result is not surprising.  

The mean difference in JND between the DG and UTI variants is \(\overline{\Delta JND} = 0.46\) and the sample variance is \(s^2 = 84.4\).
The corresponding t-statistic and p-value are 0.15 and 0.44, respectively.
Since \(p = 0.44 > 0.05\), the null hypothesis cannot be rejected. Although the mean JND of the DG variant is notably higher than that of the UTI variant (see Table \ref{tbl:JND estimates}), the within- and
between-participant variances are large. Consequently, the difference is not statistically significant according to the paired t-test.

The result that the JND in \% of the reference acceleration (which is closely related to the Weber fraction) is not affected by the motion cueing scale factor is inconsistent with the results reported by 
Baumgartner et al. \cite{Baumgartner2} and Menig \cite{Menig2025}. Both studies found that the Weber fraction increases as the reference acceleration decreases.

Baumgartner et el. \cite{Baumgartner2} proposed that the relationship between the JND in acceleration and the reference acceleration can be described by the linear model \(a_{JND} = 0.02+0.0275 a_{ref}\). 
According to this model, the predicted JND at a reference acceleration of 1.4 m/s\textsuperscript{2} is 4.2\%. If the same reference acceleration is scaled by a factor of 0.7, the predicted JND increases slightly to 4.8\%. 
Thus, the effect of scaling down the reference acceleration using a 0.7 scale factor on the Weber fraction is relatively small according to this model. 
Because only 20 trials were done with each motion cueing variant, which was too few in hindsight, the variability of the JND estimates is high. 
Because the expected effect size is relatively small and the variability in the JND estimates is high, the present study has low statistical power. We think that this is a reasonable explanation as to why we
cannot see a significant effect of the MCA scale factor on the Weber fraction.

\subsubsection{Estimating the JND and PSE}
\begin{figure}[H]
    \centering
    \begin{subfigure}[b]{0.48\textwidth}
        \centering
        \includegraphics[width=\textwidth]{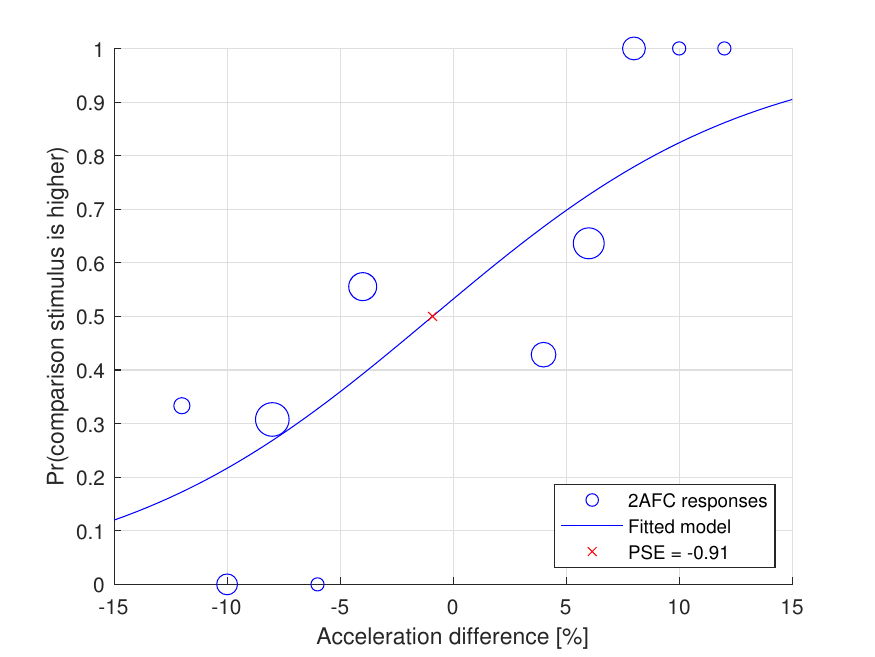}
        \caption{Driver 1}
        \label{fig:Driver 1}
    \end{subfigure}
    \begin{subfigure}[b]{0.48\textwidth}
        \centering
        \includegraphics[width=\textwidth]{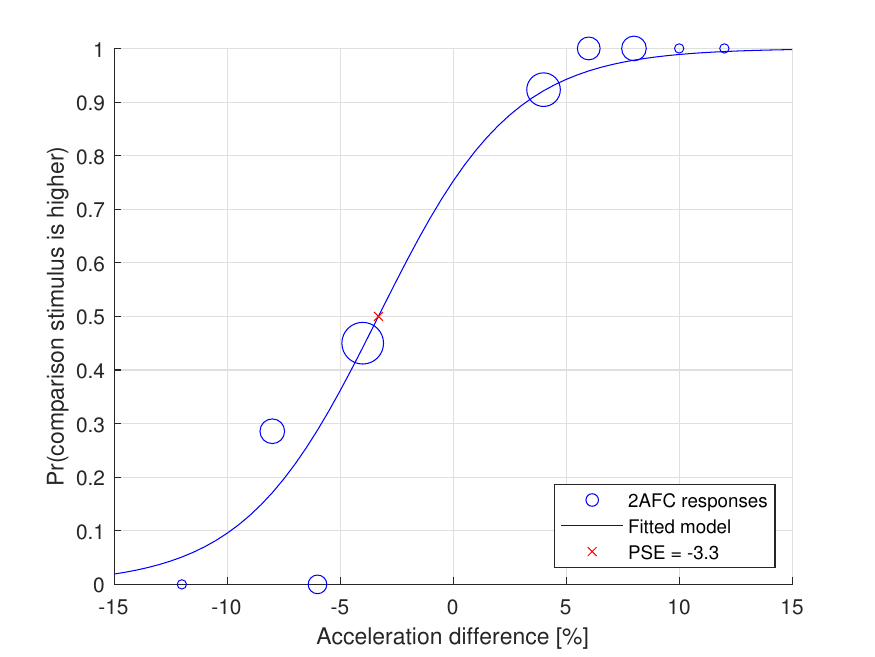}
        \caption{Driver 2}
        \label{fig:Driver 2}
    \end{subfigure}
    \begin{subfigure}[b]{0.48\textwidth}
        \centering
        \includegraphics[width=\textwidth]{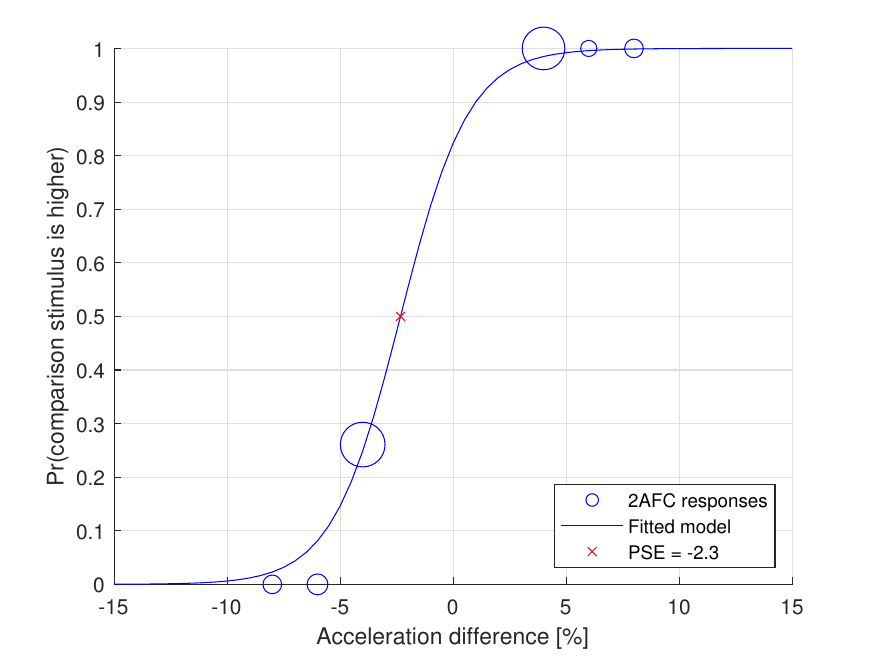}
        \caption{Driver 3}
        \label{fig:Driver 3}
    \end{subfigure}
    \begin{subfigure}[b]{0.48\textwidth}
        \centering
        \includegraphics[width=\textwidth]{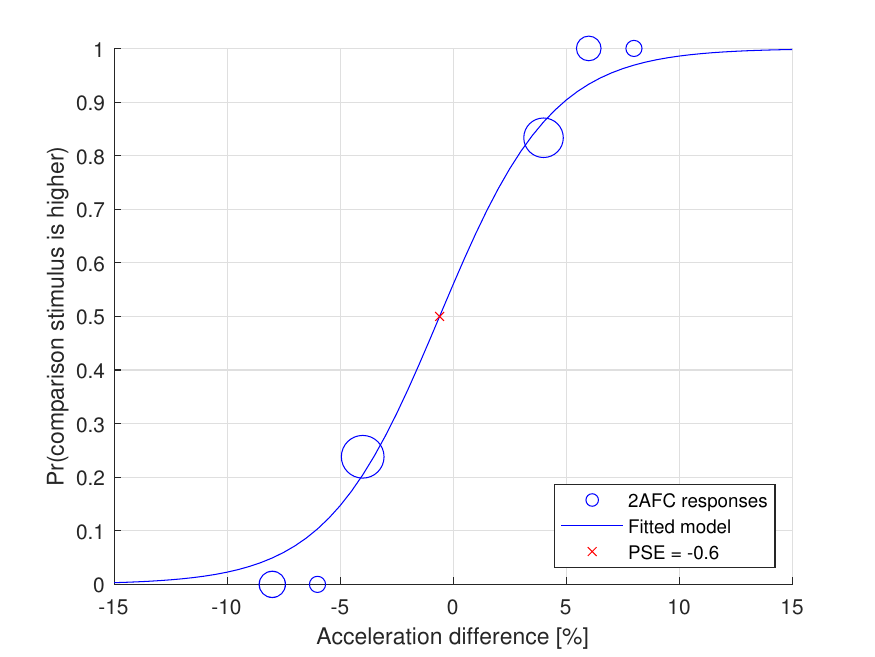}
        \caption{Driver 4}
        \label{fig:Driver 4}
    \end{subfigure}
    \caption{Fitted psychometric functions for drivers 1-4. Circle size is proportional to the number of trials conducted at each acceleration difference.}
    \label{fig:Psychomentric functions 1}
\end{figure}

\begin{figure}[H]
    \begin{subfigure}[b]{0.48\textwidth}
        \centering
        \includegraphics[width=\textwidth]{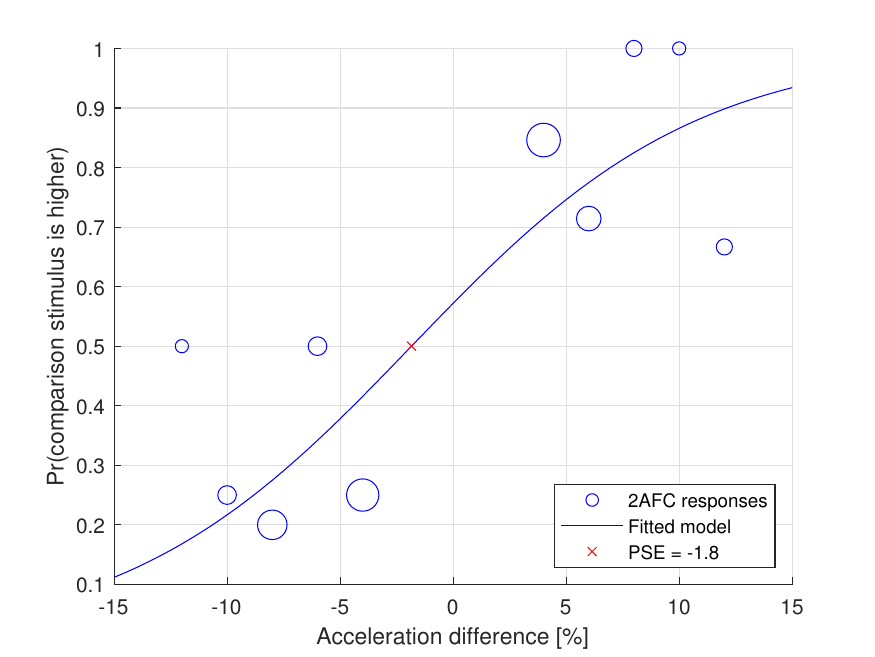}
        \caption{Driver 5}
        \label{fig:Driver 5}
    \end{subfigure}
    \begin{subfigure}[b]{0.48\textwidth}
        \centering
        \includegraphics[width=\textwidth]{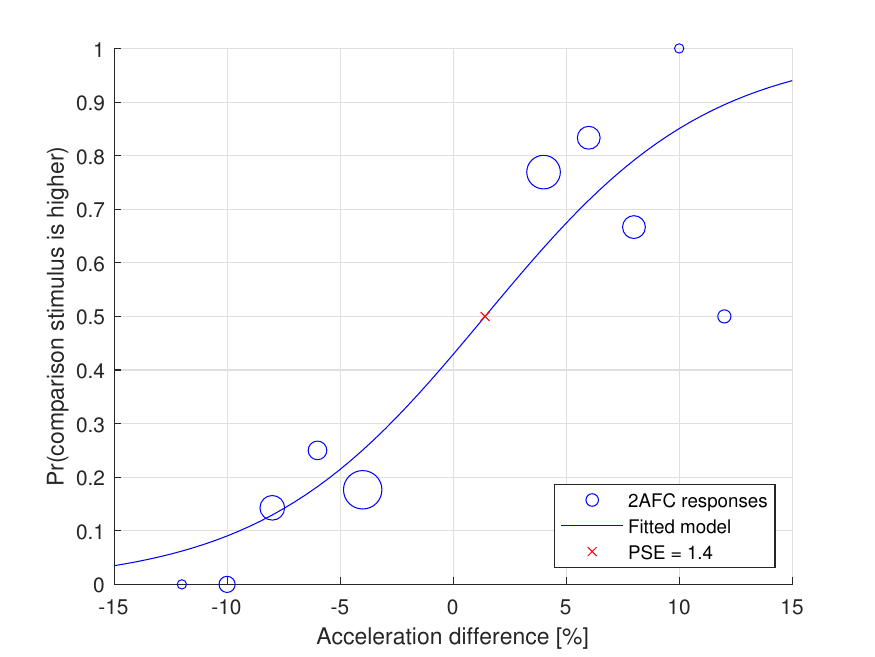}
        \caption{Driver 6}
        \label{fig:Driver 6}
    \end{subfigure}
    \begin{subfigure}[b]{0.48\textwidth}
        \centering
        \includegraphics[width=\textwidth]{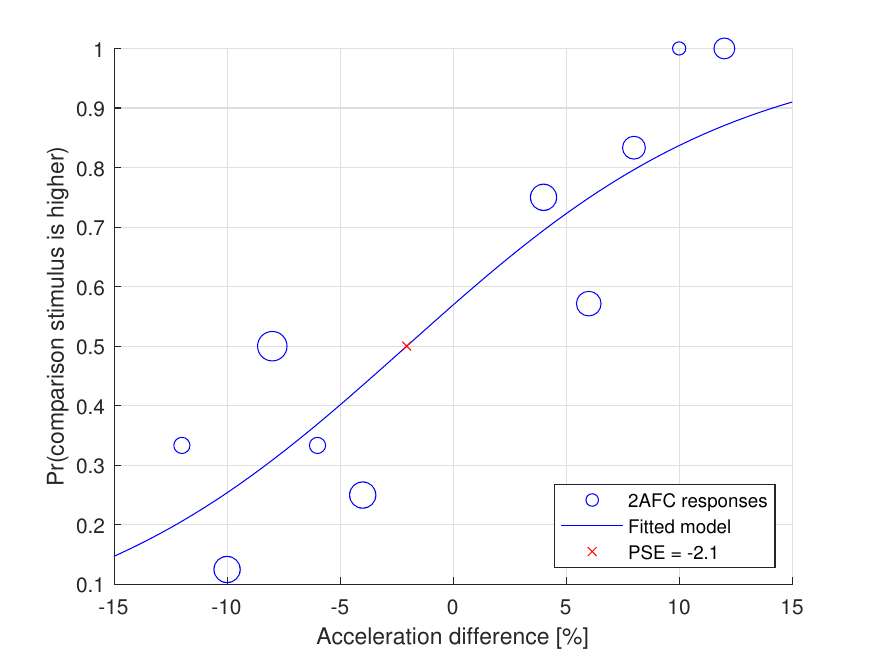}
        \caption{Driver 7}
        \label{fig:Driver 7}
    \end{subfigure}
    \begin{subfigure}[b]{0.48\textwidth}
        \centering
        \includegraphics[width=\textwidth]{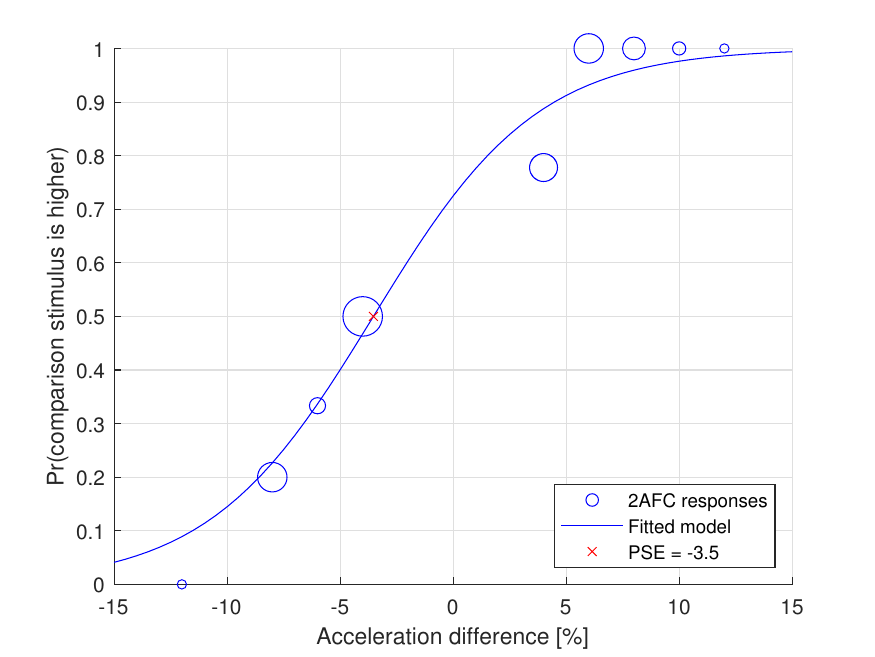}
        \caption{Driver 8}
        \label{fig:Driver 8}
    \end{subfigure}
    \\
    \begin{center}
    \begin{subfigure}[b]{0.48\textwidth}
        \centering
        \includegraphics[width=\textwidth]{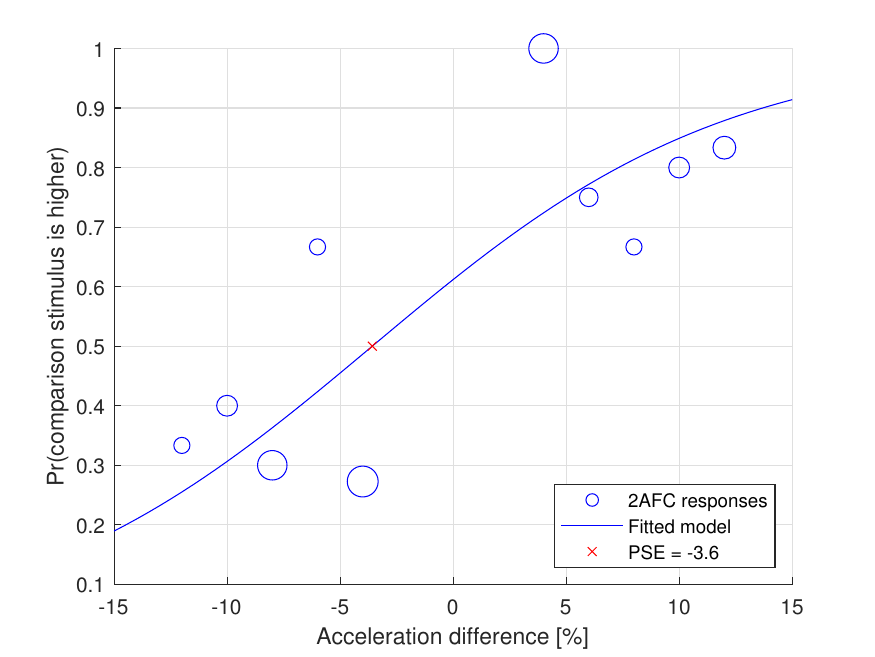}
        \caption{Driver 9}
        \label{fig:Driver 9}
    \end{subfigure}
    \end{center}
    \caption{Fitted psychometric functions for drivers 5-9. Circle size is proportional to the number of trials conducted at each acceleration difference.}
    \label{fig:Psychometric functions 2}
\end{figure}

Because the JNDs were not found to differ significantly between motion cueing variants in the previous section, the responses in the JND experiment with all three motion cueing variants were pooled together
to fit one psychometric function for each driver. The psychometric functions of the nine participants shown in Figures \ref{fig:Psychomentric functions 1} and \ref{fig:Psychometric functions 2} were
used to estimate the JNDs and PSEs in Table \ref{table:PSE and JND}. In the figures, circles represent the proportion of trials at a tested stimulus (acceleration) difference 
in which the comparison stimulus was judged to be higher than the standard.

For some participants, such as driver 1, the psychometric function is relatively flat, resulting in a large JND, whereas for others, such as driver 3, it is steep, resulting in a small JND.
For some participants, such as driver 5, the observed data is scattered and deviate considerably from the expected pattern and the fitted curve. 
This may indicate that these participants were affected by learning effects, fatigue, or lapses of attention.

The mean JND is 5.4\% which is close to the JNDs reported by Baumgartner et al. \cite{Baumgartner1} and Menig \cite{Menig2025} using similar reference accelerations.

\begin{table}[H]
    \begin{center}
        \begin{tabular}{lll}
            Driver & PSE [\%] & JND [\%] \\ 
            \hline 
            1 & -0.9 & 7.8 \\ 
            2 & -3.3 & 3.3 \\ 
            3 & -2.3 & 1.7 \\ 
            4 & -0.6 & 2.7 \\ 
            5 & -1.8 & 7.0 \\ 
            6 & 1.4 & 5.4 \\ 
            7 & -2.1 & 8.1 \\ 
            8 & -3.5 & 4.0 \\ 
            9 & -3.6 & 8.6 \\ 
            \hline 
        \end{tabular}
    \caption{PSE and JND estimates.}
    \label{table:PSE and JND}
    \end{center}
\end{table}

Interestingly, the PSE is negative for all but one participant and the mean PSE is -1.9\%. A two-tailed one-sample t-test was used to test the hypothesis that the mean PSE is non-zero.
The t-statistic is
\[
t = \frac{\overline{PSE}}{\sqrt{s^2/N}} = -3.4.
\]
This result is significant with p-value \(p = 0.009 < 0.01\) and the null hypothesis that the mean PSE is zero is rejected.
We attribute this effect to a so-called \textit{time error}, a phenomenon that has previously been observed in JND experiments \cite{Gescheider2013}. When two stimuli are presented
successively, the perceptual impression of the first stimulus fades quickly, which may lead participants to perceive the second stimulus as stronger, 
even though the two stimuli are physically identical.


\subsection{Subjective comparison}
The second MCA variant in each pair was rated on a scale from -3 to 3, where -3 indicated that the second variant was much worse than the first, and 3 indicated that it was much better.
In each pair, one variant was presented last to five participants and was presented first to the remaining four.
The number of times the second MCA variant in a pair was rated by the participants as being "much worse", "somewhat worse", "worse", "the same", "somewhat better", "better" and "much better" than the
other variant in a comparison are shown in Table \ref{table:SAs}. 
All ratings for each combination were combined into a single row, regardless of presentation order. 
For example, if a participant first tested DTI and then UTI and reported that UTI was better than DTI, this was interpreted as DTI being worse than UTI.

The overall mean ratings of the second variant in each pair, regardless of presentation order, are shown in the rightmost column of Table \ref{table:SAs}.
The mean rating for DTI relative to UTI is -0.9, indicating that participants, on average, considered UTI to be somewhat better than DTI. 
Similarly, the mean rating for DG relative to DTI is -0.8, showing than DTI was perceived as somewhat better than DG. Based on these results, one would expect UTI to also be rated better than DG. 
However, the mean rating for DG compared to UTI is -0.2, suggesting that participants perceived the UTI and DG variants as largely comparable.
This inconsistency may be due to the limited number of participants, and the results should therefore be interpreted with caution.
\begin{table}[H]
    \begin{center}
        \begin{tabular}{p{2cm} p{1cm} p{1cm} p{1cm} p{1cm} p{1cm} p{1cm} p{1cm} p{1cm}}
            \hline
            & Much worse & Worse & Some-what worse & The same & Some-what better & Better & Much better & Mean rating \\
            & -3 & -2 & -1 & 0 & 1 & 2 & 3 \\
            \hline
            UTI vs DTI & 1 & 1 & 5 & - & 2 & - & - & -0.9 \\
            DTI vs DG &	1 & 2 & 3 & 1 & 1 & 1 & - & -0.8 \\
            UTI vs DG & 2 & - & 2 & 1 & 3 & - & 1 & -0.2 \\
            \hline
        \end{tabular}
        \caption{Subjective assessments of the second MCA variant in each combination, compared with the first.}
        \label{table:SAs}
    \end{center}
\end{table}
The drivers were also given the opportunity to provide verbal assessments of the MCA variants. Participants who rated UTI as much better than other variants, 
attributed this preference to its strong and distinct acceleration cues and a higher degree of realism.  
However, one driver reported experiencing near motion sickness with the UTI motion cueing.

Many drivers were also positive about the DTI variant, describing it as natural and characterized by clear and constant acceleration. 
However, one driver noted that, compared to UTI, the DTI variant felt like playing a video game. Another driver rated DTI as worse than the DG variant 
due to a false motion cue as the platform returned to its starting position while the truck was stationary.

Drivers who rated DG lower than the other variants generally described it as synthetic and unusual, contributing to an unpleasant overall experience. 
However, one driver commented that DG was much better than UTI due to earlier response. 
Interestingly, another participant remarked that the greater cabin tilt in DG made it feel more realistic and similar to driving a tractor with a soft cab suspension.

\section{Conclusions}
This study investigated whether the JND in longitudinal acceleration in a driving simulator is lower for MCA variants tuned for launch tests 
compared to a general MCA variant, and whether the cut-off frequency of the filters affects the JND.

We did not find evidence that the MCAs tuned for the launch manoeuvre yield a lower JND than the general MCA variant. 
Instead, the JND was consistently low across all MCA variants (mean JND across variants and participants was 5.4\%), 
indicating that participants were sensitive to relatively small differences in longitudinal acceleration. Furthermore, no effect of the cut-off frequency on the JND was observed.

However, due to the limited statistical power of the study, we cannot rule out that differences between MCA variants exist. Future work with a larger sample size and improved study design is therefore needed.

Although not a primary research question, the mean point of subjective equality was -1.9\%, which was significantly different from zero (p < 0.01).
This might be due to a time-error effect, a phenomenon in which the impression of the first stimulus fades and the second is perceived as stronger despite being identical.

Finally, while no effect of motion cueing on JND was found, subjective comparisons indicated that the UTI and DTI variants, which make more extensive use of translational acceleration, were somewhat preferred by most drivers.

\bibliographystyle{elsarticle-num}
\bibliography{JND}

\end{document}